\documentclass[12pt]{article}

\usepackage{newtxtext,newtxmath}

\usepackage{graphicx}

\usepackage[letterpaper,margin=1in]{geometry}

\renewenvironment{abstract}
	{\quotation}
	{\endquotation}

\date{}

\makeatletter
\renewcommand{\fnum@figure}{\textbf{Figure \thefigure}}
\renewcommand{\fnum@table}{\textbf{Table \thetable}}
\makeatother

\usepackage{scicite}

\usepackage[hyphens]{url}
\usepackage[colorlinks=true, breaklinks=true, linkcolor=blue, citecolor=blue, urlcolor=blue]{hyperref}
\usepackage{breakurl}

\title{Electro-nuclear quantum phase transition in TmVO$_4$}

\author{
	Mark P. Zic$^{1\ast}$,
	Chao Huan$^{2}$,
	Nicolas Silva$^{2}$,
    Yuntian Li$^{3}$, \\
    Mark W. Meisel$^{2}$,
    Ian R. Fisher$^{3\ast}$ \and
	\small$^{1}$Geballe Laboratory for Advanced Materials and Department of Physics, \\ \small Stanford University, Stanford, CA 94305, USA.\and
    \small$^{2}$Department of Physics and National High Magnetic Field Laboratory High B/T Facility, \\ \small University of Florida, Gainesville, FL 32611, USA.\and
    \small$^{3}$Geballe Laboratory for Advanced Materials and Department of Applied Physics, \\ \small Stanford University, Stanford, CA 94305, USA.\and
	\small$^\ast$Corresponding authors. Emails: zic@stanford.edu and irfisher@stanford.edu\and
	}

\begin{document}

\maketitle

\title{}

\begin{abstract} \bfseries \boldmath

Hyperfine interactions couple nuclear and electronic degrees of freedom. The present work explores how hyperfine coupling within the Tm ions in TmVO$_4$ single crystals affects an electronic ferroquadrupole ordered ground state and its associated field-tuned quantum phase transition. For temperatures below the hyperfine energy scale, the nuclear moments reduce the critical field for the electronic order, resulting in a dramatic back-bending of the phase boundary delineating the ferroquadrupole order. This behavior is well described by a single-ion semiclassical mean-field model. Moreover, analysis of the effective Hamiltonian leads to a prediction of spontaneous nuclear magnetic order mediated by 4$f$ electrons, which in principle persists with the application of orthogonal antisymmetric strain, yielding a proposed electro-nuclear tetracritical point. 

\end{abstract}


\noindent

\section{Introduction}

Quantum fluctuations, which arise as a consequence of non-commuting terms in an effective Hamiltonian, can drive a continuous quantum phase transition at zero temperature between states that are characterized by different symmetries \cite{Hertz1976-classicText-quantumCriticality,Sachdev2000-classicText-quantumCriticality}. Although the quantum critical point (QCP) occurs at $T = 0$, its effects can be felt at finite temperature, for example in terms of critical scaling exponents of various observable thermodynamic quantities. In experimental condensed matter physics, quantum critical points and phenomena have been posited in a wide variety of material systems, including structural, magnetic, superconducting, and various types of charge order. Hyperfine interactions couple nuclear and electronic degrees of freedom, implying that in the race towards $T=0$, nuclear order and nuclear dynamics can also become entangled in the quantum critical state. Electro-nuclear quantum criticality has only been observed in a very small set of materials; principally LiHoF$_4$ (for which ferromagnetic order is tuned to a QCP by transverse magnetic fields) \cite{Bitko1996-LiHoF4}, PrOs$_4$Sb$_{12}$ (an antiferroquadrupolar material, also tuned by magnetic fields) \cite{Bangma2023-PrOs4Sb12-electronuclear}, and YbCu$_{4.6}$Au$_{0.4}$ (a magnetically frustrated system that is characterized by electro-nuclear fluctuations in its groundstate) \cite{Banda2023-YbCu4.6Au0.4-electronuclear}. Here, we examine the case of an electro-nuclear quantum phase transition associated with \emph{ferroquadrupolar} order. Because electronic ferroquadrupole order is necessarily accompanied by a complete softening of the crystal lattice in one symmetry channel (i.e., elastic quantum criticality \cite{Paul2017-uppercritdim}), the quantum phase transition is characterized by elastic, electronic, and nuclear quantum fluctuations, an incredible concatenation of fluctuations across enormous differences in microscopic length scales, all acting in sympathy. In addition, since the transverse field for ferroquadrupole order is a real magnetic field, the effect of hyperfine interactions is to cause an unusual back-bending of the phase boundary, which has a series of interesting consequences affecting nuclear order in this material.

TmVO$_4$ is an insulator. The $J=6$ Hund's rule ground state multiplet of the Tm ions are split by the crystal electric field (CEF), yielding a non-Kramers groundstate doublet \cite{Knoll1971-CEFspectra}. The first excited state is 77 K above the groundstate, so, at low temperatures the local electronic properties of each Tm ion can be described by a simple pseudospin representation \cite{Melcher1976-Review,Gehring1975-review}. Adding to the elegant simplicity of this material system, Thulium has a single (i.e., 100\% abundance) naturally occurring, stable isotope, $^{169}$Tm, with nuclear spin-1/2 \cite{Kondev2021-isotopesAndSpin}. Thus, at low temperatures, the individual atomic electro-nuclear states in TmVO$_4$ can be effectively described by a simple four state Hamiltonian.

Bilinear coupling between electronic quadrupoles and lattice distortions of the same symmetry provides an effective interaction between local $4f$ electric quadrupoles. In TmVO$_4$, the dominant interaction between quadrupoles with an $xy$ ($B_{2g}$) symmetry leads to ferroquadrupolar ordering with this symmetry at $T_Q = 2.15$ K via the Cooperative Jahn-Teller Effect \cite{Melcher1976-Review,Gehring1975-review}. This ordering is accompanied by a spontaneous strain of the same symmetry, leading to a coincident structural distortion from tetragonal at high temperatures to orthorhombic in the ordered state \cite{Segmuller-SponStrain}. Within the framework of the pseudospin representation, a formal mapping to the transverse field Ising model can be made \cite{Maharaj2017-TFIM} in which the effective transverse fields for the pseudospins are a magnetic field along the crystallographic $c$-direction and orthogonal antisymmetric strain (i.e., induced strains with an $x^2 - y^2$ , or $B_{1g}$, symmetry). Application of either of these effective transverse fields will suppress the long-range ferroquadrupole order \cite{Maharaj2017-TFIM}.

The field-tuned phase diagram of TmVO$_4$ was determined down to $\sim$500 mK in the 1970s via heat capacity \cite{Cooke1972-ortho} and radiofrequency susceptibility \cite{Bleaney} measurements. Those data revealed a phase boundary that appeared to closely follow expectations for a simple semiclassical mean-field single-ion treatment of the transverse field Ising model \cite{Melcher1976-Review}. More recently, magnetocaloric effect (MCE) measurements down to 670 mK were used to reveal subtle deviations from this model towards low temperatures, which at the time were attributed to the effects of quantum critical fluctuations \cite{Massat2022-PNAS}. As shown below, our present measurements reveal that this deviation, which becomes increasingly evident at yet lower temperatures, is actually associated with magnetic dipolar and hyperfine interactions. 

\section{Results}

AC susceptibility measurements were used to follow the field-tuned phase transition in TmVO$_4$ down to 10 mK, one and a half orders of magnitude lower in temperature than previously observed. The specific advantage of susceptibility measurements is that the signature of the phase transition grows progressively larger at lower temperatures (see Section \ref{supp_determining_critical_tempfield} in the Supplemental Material for calculations based on the mean-field model illustrating this), in contrast to heat capacity and MCE, which rely on entropy changes to mark the phase transition, and for which the signatures get progressively smaller in higher magnetic fields. Furthermore, AC susceptibility can be measured \emph{directly} via mutual inductance, making this signature the ideal physical quantity to follow the field-dependence of the phase transition. Two sets of experiments were performed on the same sample; one in a commercial Quantum Design dilution refrigerator insert at Stanford University (down to $\sim$110 mK), and the other using an Oxford Instruments dilution refrigerator equipped with a homemade copper nuclear demagnetization stage in the High B/T facility of the National High Magnetic Field Laboratory at the University of Florida in Gainesville (down to 10 mK). In both cases, care was taken to ensure thermalization of the sample, and checks were performed to test for hysteresis. Due to the use of a mutual inductance technique, maximizing the filling factor of the coils was prioritized when selecting the sample, so a rectangular prism (3.616 mm (along $c$) x 0.958 mm (along $a$) x 0.467 mm (along $a'$)) was used. Inhomogeneity of the internal field due to demagnetization effects within the non-ellipsoidal sample yield a slight rounding of the jump in susceptibility at the phase transition, but do not preclude obtaining a sensitive measure of the critical field/temperature for the bulk of the sample. Experiments at Stanford were repeated for multiple samples to ensure reproducibility. Details of the two experiments are provided in the Supplemental Material in Section \ref{supp_exp_details}.

\begin{figure}[t]
\centering
  \includegraphics[width=0.8\linewidth]{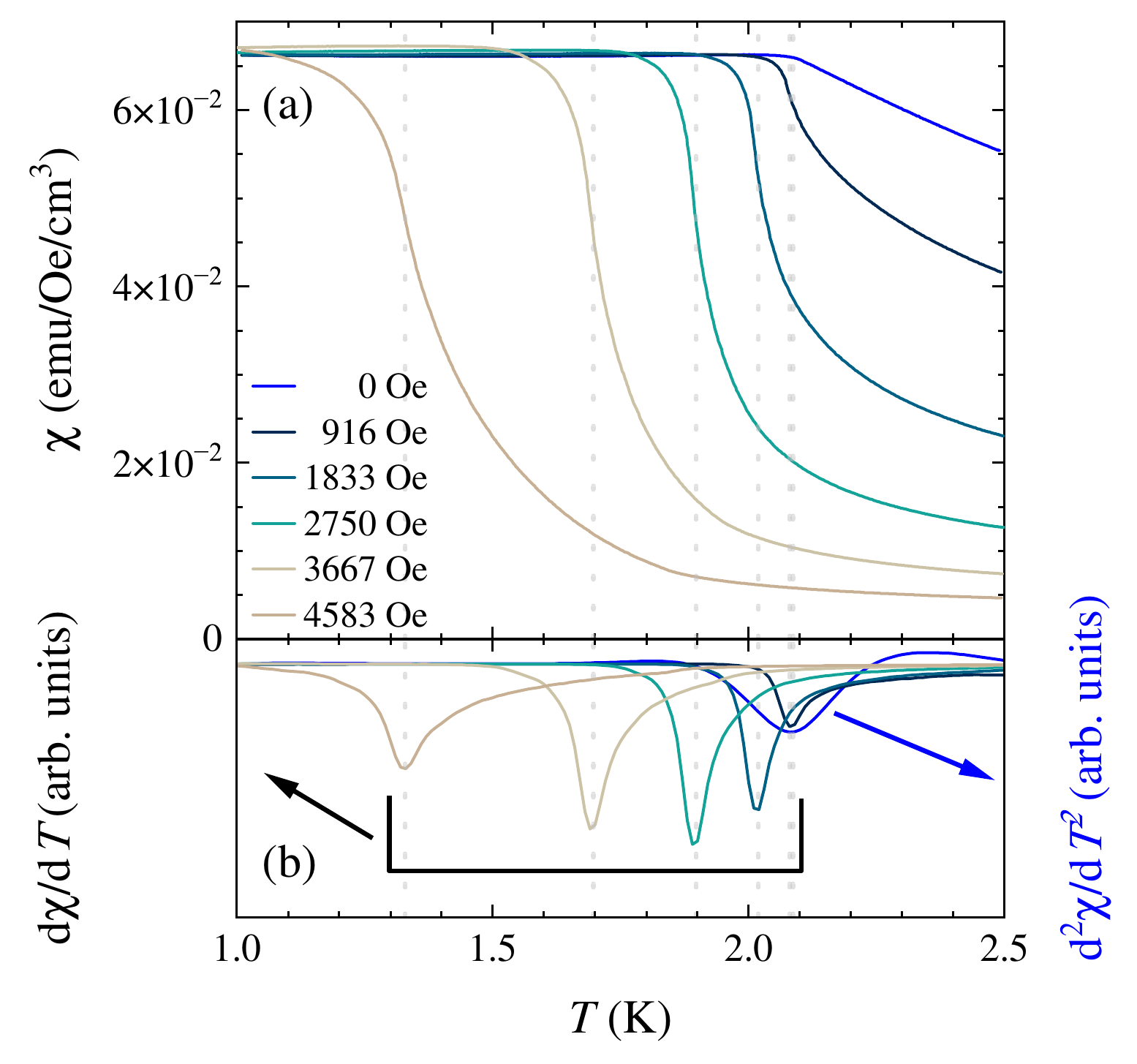}
  \caption{
\emph{Representative data showing the temperature-dependence of (a) the magnetic susceptibility, and (b) the first derivative of the magnetic susceptibility, of TmVO$_4$ for different applied magnetic fields (second derivative for $H = 0$).} The field is applied along the crystalline $c$-axis, and acts as a transverse effective field for the ferroquadrupole order. As the magnetic field is increased in magnitude, the critical temperature for the ferroquadrupolar phase transition progressively decreases. Demagnetization effects also progressively broaden the signature of the phase transition, which is identified by a sharp drop in the susceptibility, and a corresponding minimum in the first derivative. For the specific case of zero field, a minimum in the second derivative marks the phase transition (see discussion in main text). Vertical gray dotted lines mark the phase transition.}
\label{chiVsT}
\end{figure}

\begin{figure}[t]
\centering
  \includegraphics[width=0.7\linewidth]{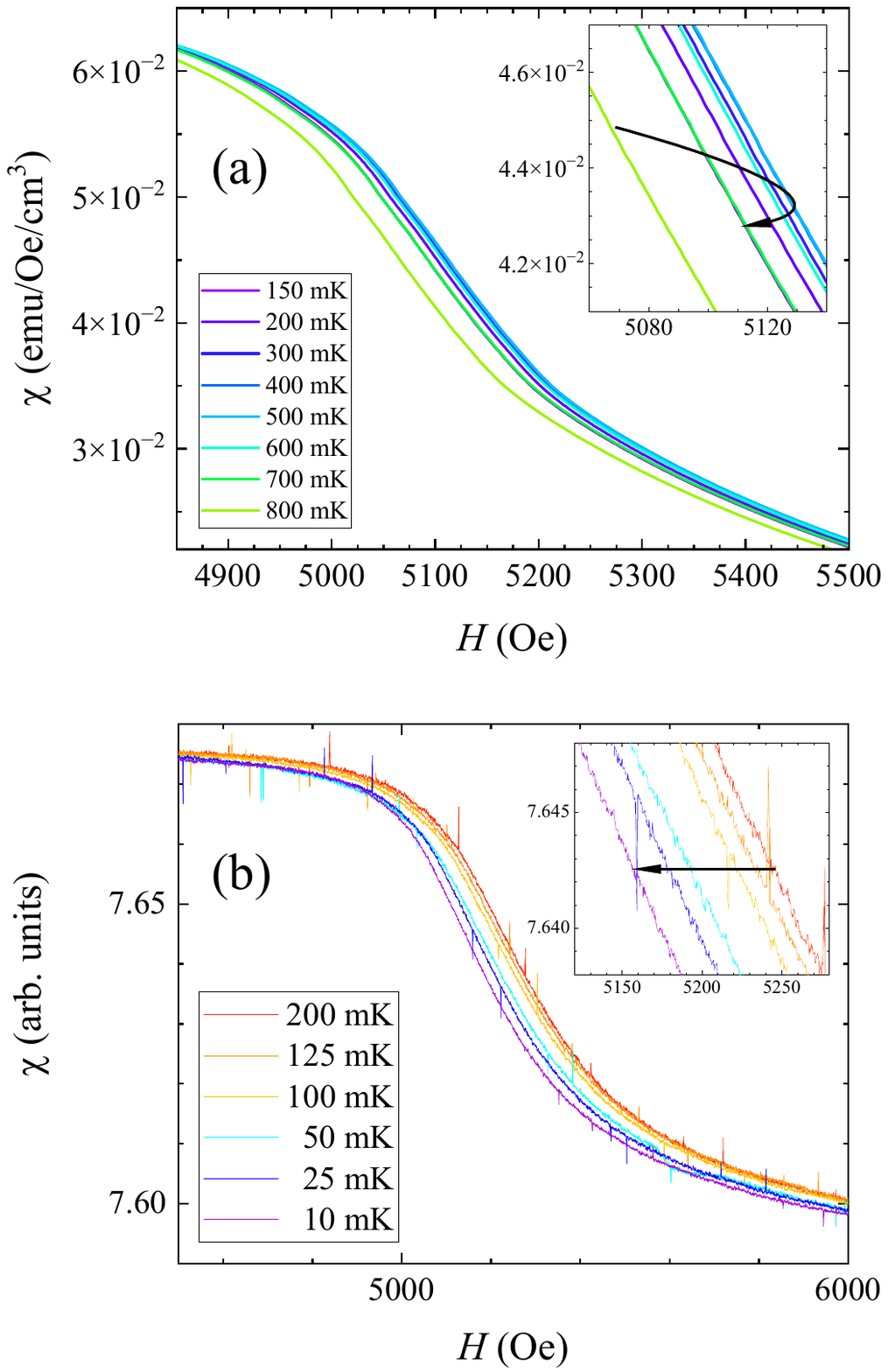}
  \caption{
\emph{Representative data showing the field-dependence of the magnetic susceptibility of TmVO$_4$ at different temperatures using (a) a dilution refrigerator at Stanford University for the temperature range from 800 mK down to 110 mK, and (b) a dilution refrigerator at the High B/T Laboratory for the temperature range from 200 mK down to 10 mK.} The critical field is marked by a sharp drop in the susceptibility (see discussion in main text). In panel (a), the critical field exhibits non-monotonic behavior, initially increasing as temperature is reduced from 800 mK, before starting to decrease as the temperature is reduced below 500 mK. In panel (b), where data were taken to even lower temperatures, the critical field monotonically decreases, but with a rate that varies with the field. Inserts to both panels show near the midpoint of the transition in each case on an expanded scale, and include arrows that point from high temperatures towards lower temperatures.}
\label{chiVsH}
\end{figure}

For small applied magnetic fields, the critical temperature does not change rapidly as the field is varied, and consequently the phase transition is most clearly seen in the magnetic susceptibility via temperature sweeps at fixed field. In zero applied field, the magnetic susceptibility exhibits a sharp kink at $T_Q \approx 2.1$ K, shown in Fig. \ref{chiVsT}(a). Below the critical temperature, the susceptibility is independent of temperature, dropping sharply as temperature is increased above $T_Q$. This behavior conforms to expectations of the single-ion mean-field model previously used to describe the properties of TmVO$_4$ \cite{Melcher1976-Review}; calculated curves are shown in Supplemental Material Section \ref{supp_determining_critical_tempfield} for comparison. As the magnetic field is progressively increased, the critical temperature decreases. The phase transition is now marked by a sharp drop in the susceptibility (see Supplemental Material Section \ref{supp_determining_critical_tempfield}), which is progressively rounded due to the change in the distribution of internal fields. Motivated by calculations of the susceptibility for the mean-field model (described in the Supplemental Material Section \ref{supp_determining_critical_tempfield}), the minimum in the first derivative of the susceptibility with respect to temperature (shown in Fig. \ref{chiVsT}(b)) is used to estimate the critical temperature for non-zero values of the field, while the second derivative is used for the case of zero field. Vertical dashed lines in Fig. \ref{chiVsT} show the critical temperature extracted from these methods. These values for the critical temperature as a function of field provide the high-temperature portion of the phase diagram.

For larger magnetic fields, the phase boundary becomes much steeper (i.e., the critical temperature varies much more rapidly with field), and consequently  field sweeps at fixed temperature are used to map the rest of the phase diagram. Representative data measured from 800 mK down to 110 mK at Stanford University are shown in Fig. \ref{chiVsH}(a); representative data measured from 200 mK down to 10 mK in the Maglab High B/T facility at the University of Florida are shown in Fig. \ref{chiVsH}(b). The phase transition is marked by a sharp downward step in the susceptibility (see Supplemental Material Section \ref{supp_determining_critical_tempfield} for mean-field calculation). This downward step is somewhat rounded in the experiment, presumably due to field inhomogeneity within the sample arising from demagnetization effects. For sufficiently slow sweeps up/down in field, the data do not exhibit hysteresis, and the phase transition appears to remain continuous. 

Insets to Fig. \ref{chiVsH} show expanded views near the midpoint of the phase transition. Inspection of these figures reveals a remarkable non-monotonic behavior of the critical field. Considering first Fig. \ref{chiVsH}(a), the critical field steadily increases as the temperature is progressively reduced from 800 mK down to 500 mK. However, below approximately 450 mK the critical field starts to decrease with further reduction in temperature (indicated by the curved arrow in the inset). Data shown in the inset to Fig. \ref{chiVsH}(b) indicate that this reduction in the critical field continues down to 10 mK.

Within the mean-field model, the phase transition is marked by a discontinuity in the susceptibility as a function of field (see Supplemental Material Section \ref{supp_determining_critical_tempfield}), and hence the minimum in the first derivative with respect to field provides the best measure of the critical field and was used to extract values from the present data. Representative data are shown in Supplemental Material Section \ref{supp_determining_critical_tempfield}.

\begin{figure}[t]
\centering
  \includegraphics[width=0.8\linewidth]{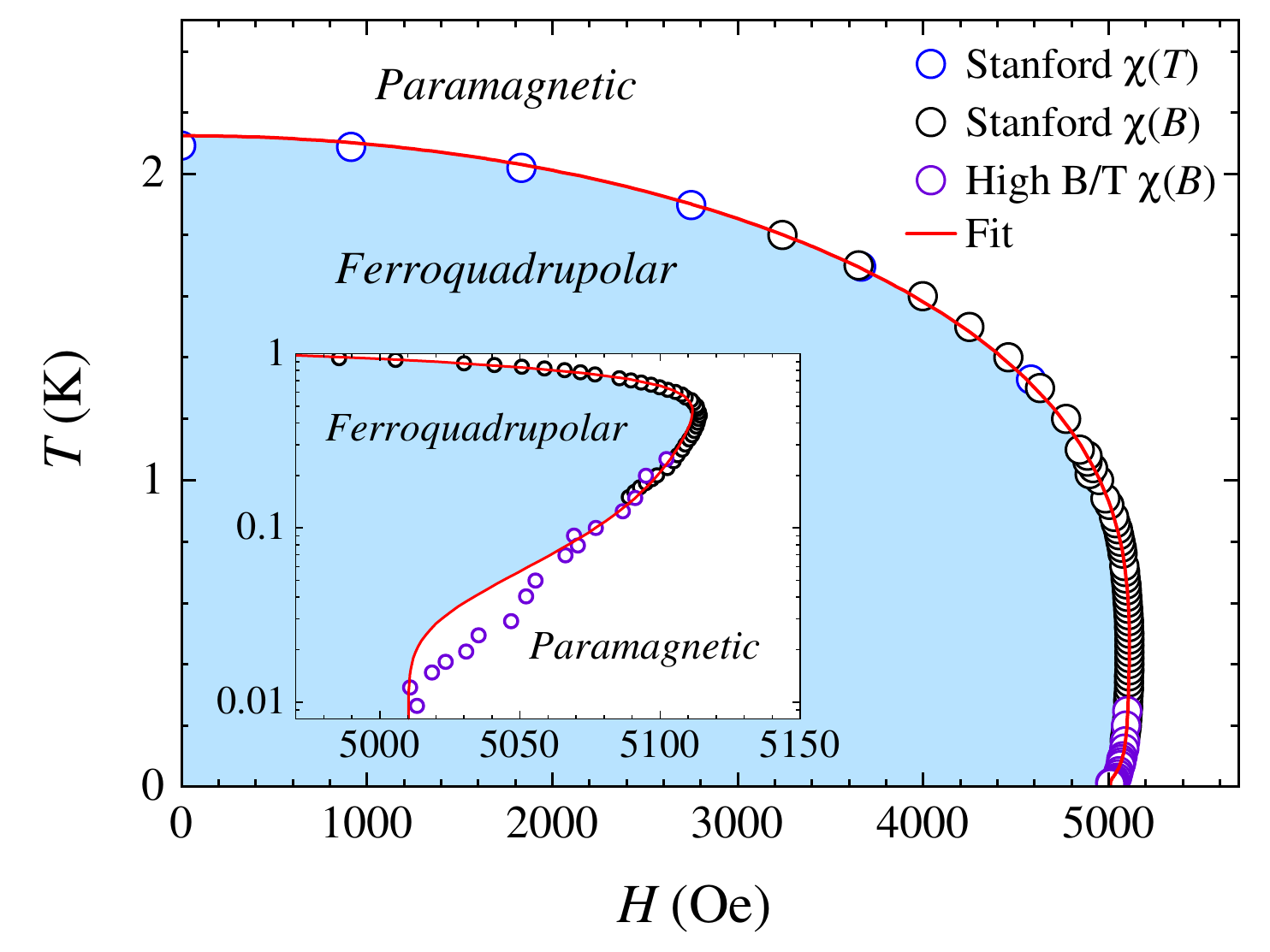}
  \caption{
\emph{Phase diagram of TmVO$_4$ as a function of applied magnetic field.} Ferroquadrupole order is shown by blue shading. The influence of the hyperfine coupling of the electrons to the Tm nucleus is especially apparent below approximately 500 mK, where back bending of the phase boundary begins. The fit to Eqn. \ref{eq_HamiltonianMeanF}, which describes the single-ion semi-classical mean-field model that includes hyperfine interactions, is shown by the red line, with fit parameters given in the main text. Inset shows the same data on a log-linear scale to more clearly reveal the remarkable back-bending of the phase boundary.}
\label{phaseDiagram}
\end{figure}

Piecing together critical temperature and critical field values from the field sweeps and temperature sweeps respectively, yields the phase diagram shown in Fig. \ref{phaseDiagram}. A remarkable back-bending of the phase boundary is clearly observed as the temperature is reduced below 500 mK. Quantitative analysis (\emph{vide infra}) indicates this behavior arises from the hyperfine interaction within the Tm ions.

\section{Mean-field description}

As mentioned previously, mean-field models have been commonly used to understand the spin interactions in TmVO$_4$, but now hyperfine interactions are explicitly included to frame our discussion of the experimental phase diagram.  Specifically, TmVO$_4$ can be described by a transverse field Ising model Hamiltonian in which the $xy$-symmetric quadrupoles are represented by spin matrices in one direction ($S^y$ here), $x^2 - y^2$-symmetric quadrupoles are represented by spin matrices in another direction ($S^x$ here), and the magnetic dipoles are represented by spin matrices in the third transverse direction ($S^z$ here) \cite{Gehring1975-review,Melcher1976-Review,Maharaj2017-TFIM}. The choice of which spin matrix to associate with which symmetry electronic state ($xy$ quadrupole, $x^2 - y^2$ quadrupole, $z$-axis magnetic dipole) is not important as long as the choice remains consistent \cite{Melcher1976-Review,Gehring1975-review}. Here, to avoid confusion, the magnetic dipole is specifically chosen to be described by the $z$-component of the pseudospin, so that the effective field for the $z$-axis component of the pseudospin corresponds to a real $z$-axis magnetic field. Hyperfine coupling links the electronic and nuclear degrees of freedom, but the strong magnetic anisotropy ($g_a \approx 0$ due to the non-Kramers nature of the CEF eigenstate) means that only the $z$ component of nuclear moment interacts with the electronic dipole moment. The Hamiltonian, in its most general form, is then:
\begin{equation} \label{eq_HamiltonianStd}
    H = -\frac{1}{2}\sum_{\langle i,j \rangle}J_{ij}^y  {S^y_i}  {S^y_j} -\Gamma^z\sum_i  {S^z_i} - \Gamma^x\sum_i {S^x_i} - \frac{1}{2} \sum_{\langle i,j \rangle} J_{ij}^z  {S^z_i}  {S^z_j} - A^z \sum_i  {I^z_i}  {S^z_i} .
\end{equation}

Here, $J^y$ is the quadrupolar coupling; the transverse effective field $\Gamma^z$ is $\frac{1}{2} g_c \mu_B B_z$, where $B_z$ is a magnetic field oriented along the crystallographic c-axis; the transverse effective field $\Gamma^x$ is $\eta_{x^2 - y^2} \varepsilon_{x^2 - y^2}$, where $\eta_{x^2 - y^2}$ is the magnetoelastic coupling for the orthogonal antisymmetric strain $\varepsilon_{x^2 - y^2}$; $J^z$ is the dipolar coupling; and $A^z$ couples the nuclear spin ($I^z$) to the electronic spin ($S^z$). It is the last term in this Hamiltonian, the hyperfine coupling to the Tm nucleus, that is imperative in understanding the low temperature behavior of TmVO$_4$ uncovered in this work.

To model the data, the Hamiltonian is simplified considerably to a mean-field, single-ion, semi-classical model. This Hamiltonian takes the form (see Supplemental Material Section \ref{supp_mean_field_theory_derivation} for more details):
\begin{equation} \label{eq_HamiltonianMeanF}
   H_{mf} = -J^y \langle {S^y} \rangle{S^y} + \big(-\Gamma^z - J^z \langle {S^z} \rangle -  A^z\langle {I^z} \rangle\big) {S^z} - \Gamma^x S^x - A^z\langle {S^z} \rangle {I^z} .
\end{equation}

Inspection of Equation 2 clearly reveals that both the dipolar interaction and the hyperfine coupling effectively strengthen the effects of the transverse field (i.e., the effective transverse field seen by the pseudospin is $(\Gamma^z+J^z \langle {S^z} \rangle +  A^z\langle {I^z} \rangle)$), which will reduce the applied field necessary to suppress the quadrupolar order. Treated at this mean-field level, however, these two terms act in slightly different ways. In the absence of hyperfine interactions, in the ferroquadrupole ordered phase, $\langle {S^z} \rangle$ is constant as a function of temperature for fixed field (a property of the transverse field Ising model). This term serves only to renormalize the effective field by a constant factor, independent of temperature, and thus does not change the functional form of the phase boundary. In contrast, the hyperfine interaction introduces an additional degree of freedom, the nuclear moment, and consequently the shape of the phase boundary is affected upon cooling through this energy scale. In this regime, the nuclear hyperfine field experienced by the electronic state progressively enhances the external field, and thus, the critical applied field that is required to destabilize the ferroquadrupole order is reduced. Ultimately, the observable consequence is that the phase diagram will exhibit ``back-bending".  

In the absence of transverse (orthogonal antisymmetric) strains (i.e., $\Gamma^x=0$), the above mean-field model contains only four free parameters: the mean-field coupling between quadrupoles, $J^y$; the g-factor $g_c$ (where $\Gamma^z$ = $\frac{1}{2} g_c \mu_B B_z$); the mean-field coupling between magnetic dipoles $J^z$, and the hyperfine scale $A^z$. In practice, all four of these quantities can be independently determined, but initially each one was treated as a fitting parameter that was varied to obtain the best fit to the observed phase boundary.

The best fit to the data (shown by the red line in Fig. \ref{phaseDiagram}) yields the parameters $J^y = 2.12 \pm 0.04$ K, $g_c = 11.44 \pm 0.24$, $J^z = 0.16 \pm 0.08$ K, and $A^z = 43.4 \pm 5.7$ mK. The coupling strength $J^y$ agrees with the zero-field critical temperature, observed here via susceptibility measurements, and in agreement with heat capacity data ($T_Q = 2.15$ K \cite{Daudin1981-specificHeat}). The value for $g_c$ is not in perfect agreement with previous optical measurements ($g_c = 10.21 \pm 0.15$ \cite{Becker1972-T-B-splitting}), but we note that it is significantly closer in value to more recent measurements based on precise ultrasound attenuation measurements in Tm$_{0.03}$Y$_{0.97}$VO$_4$, ($g_c = ~11.8$ \cite{Hollister2023-thesis}). The mean-field magnetic interaction $J^z$ closely agrees with the Weiss temperature extracted from separate longitudinal magnetic susceptibility measurements (see Supplemental Material Section \ref{supp_highT_susc_measurements}). Finally, the hyperfine coupling is within one standard deviation of the value recently obtained from ultrasound attenuation measurements ($46.5$ mK \cite{Hollister2023-thesis}). Thus, all the fit parameters are in good agreement with values that can be obtained from other, independent, measurements.

\section{Discussion}
\subsection{Electro-nuclear quantum phase transition}
Inspection of Fig. \ref{phaseDiagram} reveals that the semiclassical single-ion mean-field model, including the effects of hyperfine coupling, provides a remarkably good description of the phase boundary of TmVO$_4$ over the entire field-temperature range measured. The quality of fit, combined with the physically reasonable fit parameters obtained, implies that the origin of the observed back-bending is indeed due to hyperfine interactions.

Since the phase transition in TmVO$_4$ remains continuous down to the lowest temperatures measured, quantum critical fluctuations are anticipated upon approaching the quantum phase transition. The thermal phase transition for Ising nematic order is mean-field-like due to the long-range nematic
interaction generated by the coupling to acoustic phonons \cite{Karahasanovic_2016,Paul_2017,Qi_2009}. Since the thermal phase transition is mean-field-like, the QCP should also be. For an insulator with strong nemato-elastic coupling, theoretical work predicts a phase boundary possessing a power law dependence, $T_{Q} \sim (H_c^0-H_c)^\psi$ \cite{Massat2022-PNAS}, where $H_c^0$ is the critical field at zero temperature, and $\psi$ is given by the standard scaling relation $\psi = z/(z + d - 2)$ \cite{Lohneysen_2007}. The effective dimensionality, $d$, is 5 because the correlation length only diverges along specific directions \cite{Paul_2017}. As described in Ref. \cite{Massat2022-PNAS}, the softening of accoustic phonons, which affects the phonon dispersion near zero frequency, means that the dynamical critical exponent $z$, which characterizes temporal fluctuations of the order parameter near the QCP, will vary from a value of 1 at higher temperatures, to a value of 2 at lower temperatures. 

The electro-nuclear effects revealed in the present study significantly complicate efforts to observe this anticipated power-law behavior. Indeed, since the phase boundary can be adequately described by the semi-classical result, which neglects quantum fluctuations of the order parameter, one may conclude either (1) that the quantum critical dynamics do not manifest until significantly below the lowest temperature measured (10 mK); or (2) that the progressive effect of hyperfine interactions (and possibly also dipolar and exchange interactions if these act in a non-mean-field way) dwarfs any more subtle effects associated with growing quantum critical fluctuations, at least in the range of temperatures considered here. Either way, experiments to yet lower temperatures will be required, in order to freeze out any temperature dependence associated with the hyperfine interaction. The mean-field model developed here indicates that below approximately 10 mK, the phase boundary will cease back-bending, providing an upper bound below which subsequent measurements should test for the theoretically predicted power law behavior of the phase boundary. This is unfortunately true of other physical quantities for which quantum fluctuations are anticipated to yield power laws, including the longitudinal susceptibility, motivating a concerted effort to study this material in the sub mK regime.  

Fits to the mean-field model do reveal small systematic deviations, but one cannot conclude with any confidence that these are significant. Fitting our data over a similarly small temperature range as was done in the earlier study ($T_Q$ down to $\sim$1.5 K) \cite{Massat2022-PNAS}, also yields an extrapolated fit that overshoots the data, possibly implying that the effects of quantum fluctuations begin to be felt before the hyperfine interaction yields the more dramatic back-bending of the phase boundary. However, given the overall quality of fit to the semiclassical model, it seems more likely that the apparent power law that was previously observed between $\sim$0.7 and 1.4 K \cite{Massat2022-PNAS} reflects instead a subtle progressive change in slope of the measured phase boundary, casting doubt on the earlier suggestion that this was indirect evidence of quantum criticality. Similarly, inspection of the inset of Fig. \ref{phaseDiagram} at low temperatures, below approximately 0.5 K, reveals a systematic deviation of the data from the best fit. In this regime, the data have an apparent linear variation on the log-linear plot, implying power law behavior in this regime. However, we caution against interpreting this behavior as evidence for quantum criticality until data are available to a significantly lower temperature, where the progressive effects of the hyperfine interaction are frozen out.

With our observations, TmVO$_4$ joins only a handful of other materials for which nuclear interactions have been shown to play an important role in the electronic order and/or quantum phase transition, including LiHoF$_4$ \cite{Bitko1996-LiHoF4}, PrOs$_4$Sb$_{12}$ \cite{Bangma2023-PrOs4Sb12-electronuclear}, YbCu$_{4.6}$Au$_{0.4}$ \cite{Banda2023-YbCu4.6Au0.4-electronuclear}, YbRh$_2$Si$_2$ \cite{Knapp2023-YbRh2Si2-electronuclear,Nguyen2021-YbRu2Si2-electronuclear,Schuberth2016-YbRh2Si2-electronuclear}, PrCu$_2$ \cite{Andres1973-PrCu2-electronuclear}, and Pr$_3$Pd$_{20}$Ge$_6$ \cite{Iwakami2014-Pr3Pd20Si6-electronuclear,Steinke2013-Pr3Pd20Si6-electronuclear}. Of these materials, LiHoF$_4$ provides a particularly insightful point of comparison for TmVO$_4$. Both materials are tetragonal, though with different crystal structures. The magnetic ion in both cases has the same symmetry non-Kramers CEF doublet ground state (i.e., belongs to the same irreducible representation of the point group), which can harbor a $c$-axis magnetic dipole and the two in-plane quadrupoles (with $x^2-y^2$ and $xy$ symmetry). As described above, TmVO$_4$ is an Ising quadrupolar system, with long-range quadrupolar interactions. LiHoF$_4$, however, orders via its dipole moment, and is an Ising ferromagnet with long-range dipolar interactions. The thermal phase transitions in both materials are of interest: the Ising model with long-range strain interactions has an upper critical dimension $d^+ = 2$ \cite{Paul_2017}, so the phase transition in TmVO$_4$ is mean-field-like \cite{Cooke1972-ortho}, whereas $d^+ = 3$ for the Ising model with long range dipolar interactions, such that LiHoF$_4$ exhibits marginal dimensionality \cite{Griffin_1980}. Both systems can be mapped onto a variant of the transverse field Ising model, with an in-plane magnetic field ($H_x^2$) acting as the transverse field for LiHoF$_4$ \cite{Bitko1996-LiHoF4}, and a c-axis magnetic field ($H_z$) acting as the transverse field for TmVO$_4$. Both nuclei, Ho and Tm are 100\% abundant with a single isotope, though they differ in the nuclear spin, with Tm having the simplest case of $I=1/2$. Hyperfine interactions in LiHoF$_4$ serve to strengthen the ferromagnetic order, such that the field-tuned phase boundary extends further than if there were no hyperfine interactions \cite{Bitko1996-LiHoF4}; whereas the hyperfine interactions in TmVO$_4$ enhance the effect of the transverse field, leading to the `back-bending' of the phase boundary shown in Fig. \ref{phaseDiagram}. The two materials are compared in Table 1; both are model systems for studying thermal and quantum phase transitions, exhibiting sufficient complexity to be interesting, but sufficient simplicity to be tractable (from both experimental and theoretical perspectives). \\

\begin{table}[t]
\centering
\begin{tabular}{c|c|c|c|c|c|c}
  \hline
  Material & Active Ion & Ordering & $T_c$ & $I$ & $J$ & $N_{H_{MF}}$  \\ 
  \hline
  TmVO$_4$ & Tm$^{3+}$ & Ising ferroquadrupolar & 2.15 & $1/2$ & 6 & 4\\ 
    LiHoF$_4$ & Ho$^{3+}$ & Ising ferromagnet & 1.53 & $7/2$ & 8 & 136 \\ 
\end{tabular}
\caption{\emph{Comparison between TmVO$_4$ and LiHoF$_4$.} The active ions, ordering, transition temperature, nuclear spin $I$, electronic total angular momentum $J$, and the size of the mean-field matrix $N_{H_{mf}}$ are listed for both materials. Note that, given the nature of the crystal field eigenstates and the transverse fields for each material, the Hamiltonian used to model TmVO$_4$ is of much lower rank, greatly adding to its appeal as a model system to study quantum phase transitions.}
\label{LiHoF4_TmVO4_comparison}
\end{table}

\subsection{Implications for nuclear order}

Nuclear moments are considerably smaller than electronic ones, with the ratio of the nuclear and Bohr magnetons being $\mu_N/\mu_B=m_e/m_p$ where $m_e$ and $m_p$ are the mass of the electron and proton respectively. Nevertheless, nuclear order can occur at a substantially elevated temperature relative to that which would be anticipated if the nuclear moments interact solely through their dipolar fields. When electronic magnetic order exists, hyperfine interactions necessarily split any nuclear degeneracy and lead to an induced nuclear moment on these ions, precluding spontaneous nuclear magnetic order. However, for van Vleck materials, which have a singlet CEF ground state but an enhanced paramagnetic susceptibility due to the presence of low-lying CEF excited states, the nuclear degeneracy can be lifted through spontaneous nuclear magnetic order, with the nuclear moments coupled via the electronic system \cite{Ishii_2004}. In these systems, the hyperfine interaction enables the nuclear moment to co-opt some small amount of $4f$ dipole moment on each ion, leading to an enhanced effective nuclear moment, $\mu=g_N\mu_N(1+K)\langle I \rangle $, where $g_N$ is the nuclear g-factor and $K$ is an element and material specific enhancement factor. Examples include the intermetallic comounds PrCu$_2$ \cite{Andres_1972}, PrCu$_6$ \cite{Babcock_1979}, PrNi$_5$ \cite{Kubota_1980}, and the insulator HoVO$_4$ \cite{Suzuki_1978} (which belongs to the same crystal structure as TmVO$_4$). The absence of magnetic order in TmVO$_4$ (which, courtesy the quadrupolar order, renders itself a van Vleck magnet below $T_Q$) makes this material a natural candidate for nuclear magnetic order. The key difference to all other known van Vleck nuclear magnets is that the splitting of the CEF states in TmVO$_4$ is not caused by the crystal symmetry, but rather is induced by the quadrupolar phase transition, and therefore can be modulated by the material in order to minimize the free energy. In the pseudospin language, the electronic pseudospin can spontaneously rotate on its Bloch sphere, moving away from the pure quadrupole `direction' in order to acquire some small dipole moment. This leads us to propose a novel interplay between the nuclear and quadrupolar order in this material at very low, but hopefully accessible, temperatures.  

In the case of TmVO$_4$, the enhancement factor $(1+K)$ for the Tm ion was previously estimated to have a value of 774 \cite{Bleaney,Suzuki_1980}, indicating the possibility of nuclear magnetic order at an elevated temperature. Bleaney and Wells estimated a critical temperature based on calculations of the nearest neighbor dipole-dipole interactions, finding that ferromagnetic order is favored with an estimated critical temperature of 0.25 mK \cite{Bleaney}. They noted that a small amount of superexchange interaction could favor an antiferromagnetic state, estimating in that case a critical temperature of 0.28 mK based on estimates of the induced field on the V nuclei \cite{Bleaney}. Including exchange interactions within the mean-field model developed herein, and using the same values for the fit parameters described above, a non-zero expectation value of the nuclear order parameter at zero temperature, and, for zero applied magnetic field, a phase transition to a ferromagnetically ordered state deep within the ferroquadrupole state, is also found. Based solely on the fit parameters obtained from the phase diagram, which indicate a dominant ferromagnetic coupling between Tm dipole moments, we obtain a somewhat higher predicted mean-field critical temperature than Bleaney and Wells, $T_{mf} = (A^z)^2/(J^y - J^z) = 0.96\ \mathrm{mK}$ (see Supplemental Material Section \ref{supp_nucl_ordering_temp_derivation}). Despite these predictions, nuclear demagnetization experiments performed on TmVO$_4$ by Suzuki \emph{et al.} in 1980 found no evidence for magnetic order down to 0.1 mK \cite{Suzuki_1980}, indicating the presence of frustration of one sort or another, potentially reflecting the competing effects of superexchange and dipolar interactions. In the following analysis, we use the calculated mean-field critical temperature $T_{mf}$ in order to make a couple of important symmetry-related points (i.e., conclusions that do not depend on microscopic details, but only on the symmetry of the ordered state), while noting that the actual critical temperature is clearly somewhat lower in temperature.

\begin{figure}
\centering
  \includegraphics[width=0.90\linewidth]{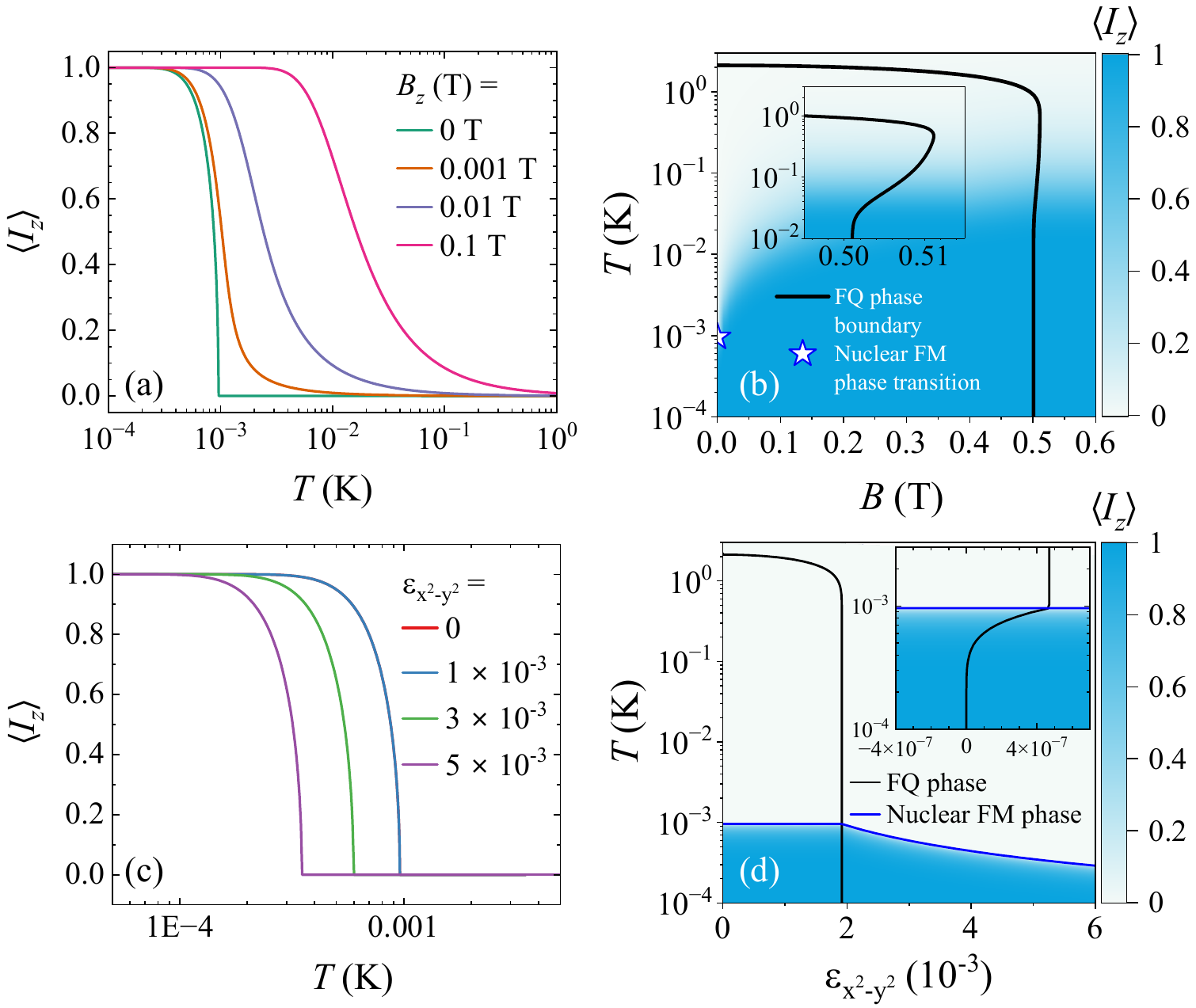}
  \caption{
\emph{Predictions for nuclear order in TmVO$_4$ based on the mean-field model, as a function of the two transverse effective fields.} The temperature-dependence of the nuclear order parameter $\langle I_z \rangle$ is shown in panels (a) and (c) for different values of the applied magnetic field and applied strain $\epsilon_{x^2-y^2}$, respectively. The associated phase diagrams are shown in panels (b) and (d), together with a heat map (blue color scale) of the nuclear order parameter. In both figures, the ferroquadrupolar phase boundary is depicted by a solid black line. In panel (b), the nuclear ferromagnetic phase transition predicted by this mean-field model (see discussion in main text) occurs at approximately 0.96 mK and is marked by a blue star. The phase transition only occurs in zero field, and is smeared out for non-zero values of $B$. The expectation value of the nuclear spin operator begins to grow at higher temperatures as field is progressively increased, directly affecting the shape of the phase boundary; the temperature at which the phase boundary `back-bends' corresponds to the temperature at which the nuclear moment starts to grow rapidly (shown on an expanded scale in the inset). In panel (d), the solid blue line corresponds to the nuclear ferromagnetic phase transition, which persists as a function of strain. Inset shows an expanded scale close to the predicted tetracritical point (see discussion in main text). }
\label{calculatedPhaseDiagram}
\end{figure}

First, if the nuclear order is indeed ferromagnetic (as the observed Weiss temperature implies) then in the presence of a non-zero magnetic field there will be no spontaneous phase transition since time-reversal symmetry is already broken by the magnetic field (Fig. \ref{calculatedPhaseDiagram}(a)). Consequently, the complete phase diagram for TmVO$_4$ as a function of magnetic field would comprise the ferroquadrupole ordered state and a single critical point for the ferromagnetic electro-nuclear order. This is illustrated in Fig. \ref{calculatedPhaseDiagram}(b), in which the predicted mean-field critical point for the ferromagnetic order $T_{mf}$ is shown by a star. For comparison, a heat map of $\langle I^z \rangle$ is overlaid, demonstrating that a non-zero magnetic field smears out the phase transition and graphically illustrating how the observed `back-bending' of the phase diagram coincides with the growing nuclear moment as temperature is reduced below approximately 500 mK.

While the field-tuned phase diagram is hardly modified by the nuclear order, having only a single critical point in zero magnetic field, the predicted strain-tuned phase diagram is significantly richer. As mentioned above, orthogonal antisymmetric strain $\varepsilon_{x^2-y^2} = \varepsilon_{xx}-\varepsilon_{yy}$ is also a transverse effective field for the $xy$ symmetry quadrupole order. Since strain does not break time reversal symmetry, spontaneous nuclear ferromagnetic order is still allowed, and hence the electro-nuclear order persists across the entire phase diagram. Figures \ref{calculatedPhaseDiagram}(c) and \ref{calculatedPhaseDiagram}(d) show the results of the mean-field model for $\langle I^z \rangle$ and the associated phase diagram. The mean-field theory predicts an electro-nuclear tetracritical point, in which the purely electronic quadrupolar phase boundary crosses the electro-nuclear ferromagnetic phase boundary. As temperature is decreased below the tetracritical point, the quadrupolar phase boundary exhibits a subtle back-bending (shown on an expanded scale in the inset to Fig. \ref{calculatedPhaseDiagram}(d)). In the ferromagnetic phase, the pseudo-spin describing the electronic quadrupole rotates slightly to acquire some small amount of magnetic dipole moment. In so doing, the critical strain to suppress the quadrupole order is reduced, leading to the back-bending. Below this temperature, the two co-existing orders are remarkably intertwined, since the electronic states contribute to both the electronic quadrupole order and the electro-nuclear ferromagnetism. Finally, the ferromagnetic order is predicted to persist for strains beyond the critical strain for the quadrupole order (i.e., for strains beyond approximately $2\cdot10^{-3}$), but the critical temperature grows progressively smaller with increasing strain as the strain-induced splitting of the CEF doublet grows progressively larger. 

Although earlier experiments indicate that the nuclear order in TmVO$_4$ occurs at a lower temperature than the mean-field estimate calculated in this work, nevertheless it is inevitable that the nuclear moments will eventually order, and the large enhancement factor $K$ implies a relatively high critical temperature. Our theoretical predictions motivate performing the challenging experiments that will be necessary to explore the remarkable phase diagram predicted in Fig. \ref{calculatedPhaseDiagram}(d).

\subsection{Implications for cooling via nuclear demagnetization}

The back-bending of the phase boundary in TmVO$_4$ is superficially reminiscent of the pressure-tuned phase diagram of $^3$He. However, the phase transition in TmVO$_4$ is continuous, distinct from the case of $^3$He, and hence is not governed by the same Clausius-Clapeyron relation that accounts for Pomeranchuk cooling. A cooling effect is nevertheless still predicted for TmVO$_4$, but it arises from nuclear demagnetization, and changes in the magnitude of the cooling are governed by an Ehrenfest relation.

\begin{figure}[h]
\centering
  \includegraphics[width=1\linewidth]{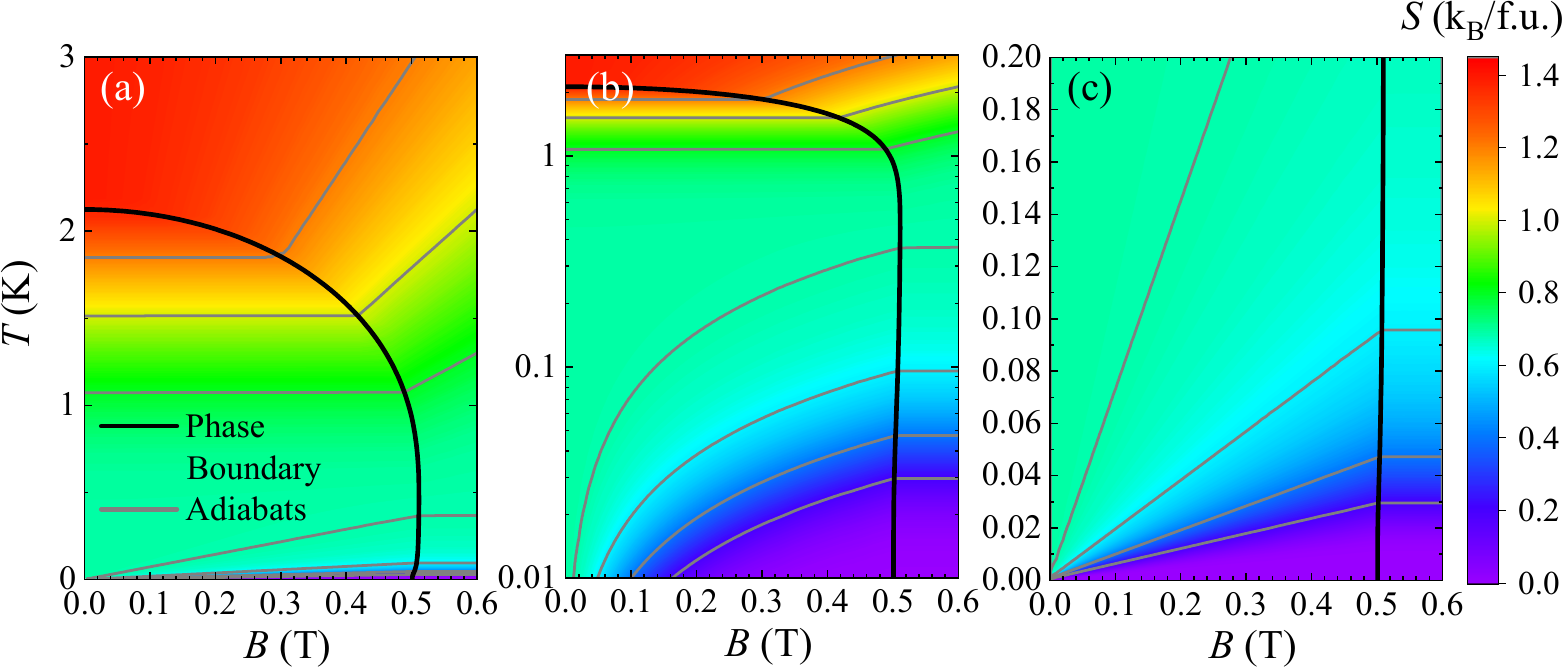}
  \caption{\emph{Calculated entropy of TmVO$_4$ as a function of temperature and magnetic field, illustrating regimes where cooling from electronic and nuclear demagnetization dominate respectively.} The entropy is expressed per formula unit (f.u.; i.e., per Tm ion) and is displayed as a color scale on (a) a linear-linear scale over the entire region of interest; (b) a semi-log scale (temperature in log, field in linear) over the same region; and (c) over a smaller region for low temperatures (0 K to 0.2 K) on a linear-linear scale. The phase boundary is shown by a black solid line. Gray lines show representative constant entropy contours (isentropes). Changes in field and temperature lead to fully reversible changes in entropy, thus isentropes are also adiabats. Adiabatic demagnetization would correspond to following any of the gray lines from high field towards zero field. For temperatures above approximately 700 mK, adiabatic demagnetization results in the largest cooling effect outside the ferroquadrupole ordered state, with negligible effect inside the phase boundary (seen most clearly in the top half of panel (a)). For temperatures below this, the reverse is true: the largest cooling effect is found within the ferroquadrupolar phase, with negligible cooling outside (seen most clearly as the change in slope of isentropes upon crossing the phase boundary in the lower half of panel (b) and in panel (c).)  The cross over between these two regimes occurs as the expectation value of the nuclear order parameter grows (i.e., when the back-bending of the phase diagram begins, which coincides with the growing expectation value of $I_z$ in Fig. \ref{calculatedPhaseDiagram}).  }
\label{fullEntropyLandscapes}
\end{figure}

The mean-field model that we use to fit the data can also be used to calculate the entropy across the H-T plane. At these low temperatures, the phonon contribution to the entropy is negligible, though for completeness is still included in the calculations. For temperatures above where the electro-nuclear coupling has little effect on the phase boundary (i.e., above approximately 700 mK), the entropy is nearly constant as a function of field inside the ordered state. This is a standard result of the transverse field Ising model; neglecting the nuclear states, the separation in energy of the two atomic states does not change with field, leading to a constant entropy. However, below 700 mK, in the regime where the phase boundary exhibits the back-bending, the entropy depends on the strength of the magnetic field, rising with increasing field inside the ordered state (see Fig. \ref{fullEntropyLandscapes}). In this regime, the contribution to the field-dependence of the entropy arises from the splitting of the electro-nuclear doublet, which is zero in zero field (for temperatures above the predicted nuclear ferromagnetic order) and rises to progressively larger values as the field is increased, as the amount of admixed $4f$ moment progressively increases. For fields beyond the critical field, the entropy continues to rise with increasing field, but at a much lower rate (appearing almost flat in Fig. \ref{fullEntropyLandscapes} over the small field range plotted). In this regime, the electro-nuclear doublet is progressively split by the applied field, but the amount of $4f$ moment that is admixed is no longer determined by the suppression of the quadrupole order and hence varies much more slowly with field. 

Since the ferroquadrupole phase transition is continuous, an Ehrenfest relation (see Supplemental Material Section \ref{supp_ehrenfest_relation}) relates the slope of the phase boundary $dT_Q/dH$ to the jump in the magnetocaloric effect at constant entropy $\Delta(\partial T/\partial H)_S = (\partial T/\partial H)_S^{T_Q^-} - (\partial T/\partial H)_S^{T_Q^+}$ and the jump in the heat capacity at constant field $\Delta C_H = C_H^{T_Q^-}-C_H^{T_Q^+}$ at $T_Q$ (where values are measured just below and just above $T_Q$, defining $T_Q^-$ and $T_Q^+$): 

\begin{equation}
    \frac{d T_Q}{dH} = \frac{\Delta (C_H (\frac{\partial T}{\partial H})_S)}{\Delta C_H}.
\end{equation}

The back-folding of the phase transition below approximately 500 mK marks a change in sign of $dT_Q/dH$ and delineates the regime in which cooling from adiabatic demagnetization is stronger inside the ferroquadrupolar state than outside it (i.e., $\Delta (\frac{\partial T}{\partial H})_S >0$). This regime is the one governed by nuclear demagnetization. The regime above approximately 500 mK is primarily electronic, with cooling governed by adiabatic demagnetization of the electronic energy levels, which is only operative outside the ferroquadrupole ordered state. Following the adiabats (gray lines) in Fig. \ref{fullEntropyLandscapes} clearly illustrates these two regimes, and the cross over between them. In particular, cooling via electronic demagnetization is largest in the paramagnetic phase for temperatures above 700 mK, while cooling via nuclear demagnetization is largest inside the ferroquadrupolar phase for temperatures below this value.

TmVO$_4$ was previously suggested to be a candidate for use in a nuclear demagnetization refrigerator \cite{Bleaney}, made plausible by the relatively low critical temperature for nuclear magnetic order and short spin-lattice relaxation time \cite{Suzuki_1980,Suzuki_1981}. By observing the back-folded ferroquadrupole phase boundary, directly revealing the effect of hyperfine interactions on the ferroquadrupole order, our work also reveals the necessary change in slope of $(\partial T/\partial H)_S$ upon crossing the phase boundary. In contrast to the Pomeranchuk effect, the cooling effect in TmVO$_4$ would arise simply from adiabatic demagnetization (following the gray lines from high to low field in Fig. \ref{fullEntropyLandscapes}). The back-bending of the phase diagram reflects the underlying electro-nuclear splitting, but is not a driving force for the cooling effect \emph{per se}. Rather, both effects originate from the same hyperfine interaction.\\

\section{Acknowledgments}
We thank N. Curro, R. Fernandes, P. Hollister (who first pointed out the possible effects of hyperfine interactions on the ground state of TmVO$_4$), J. Hu, S. A. Kivelson, D. Petruzzi, and B. J. Ramshaw for helpful conversations. Crystal growth and low temperature susceptibility measurements performed at Stanford were supported by The Airforce Office of Scientific Research under award FA9550-20-1-0252 and award number FA9550-24-1-0357; data taken at Stanford were obtained using a cryostat acquired with award FA9550-22-1-0084. The work at the University of Florida and High B/T Facility, as a part of the National High Magnetic Field Laboratory (Maglab), was supported by the National Science Foundation through NSF/DMR-2128556 and the state of Florida. M. P. Zic was partially supported by a National Science Foundation Graduate Research Fellowship under award DGE-1656518.

\section{Data Availability}
Data used in the preparation of this manuscript can be found in the Stanford Digital Respository at \href{https://purl.stanford.edu/ky369kj8044}{https://purl.stanford.edu/ky369kj8044}.

\clearpage 

%
\bibliography{TmVO4_phaseDiagram} 
\bibliographystyle{sciencemag}

\newpage

\renewcommand{\thesection}{S\arabic{section}}
\setcounter{section}{0}

\section{Experimental methods}

\subsection{Crystal growth} \label{supp_growth}
Measurements were performed on a sample of TmVO$_4$ shaped as a rectangular prism with dimensions 3.616 mm (along the crystallographic $c$-axis) by 0.958 mm (along the crystallographic $a$-axis) by 0.467 mm (along the orthogonal $a$-axis). The sample was grown using a flux growth method \cite{Feigelson,Smith,Massat2022-PNAS} and subsequently cleaned, cut with a wire saw, and polished using fine sandpaper.

\subsection{AC susceptibility measurements} \label{supp_exp_details}
AC magnetic susceptibility measurements were performed at both Stanford University and the National High Magnetic Field Laboratory (MagLab) High B/T Facility at the University of Florida. Measurements at Stanford University were performed using a commercial Quantum Design DynaCool with a dilution refrigerator insert, using a field amplitude of 3 Oe and a frequency of 643 Hz. No frequency dependence as observed in the range tested (approximately 100 to 1000 Hz). Measurements at the MagLab High B/T Facility were performed using a homemade ultralow temperature AC susceptometer, which allowed the sample to be entirely immersed in liquid $^3$He thermalized by a silver sinter heat exchanger, mounted on the Bay 2 instrument, using a field amplitude of 0.01 Oe and a frequency of 106.7 Hz. The lower frequency and amplitude used at the MagLab High B/T Facility were chosen to reduce heating effects at low temperatures. In both cases (i.e., measurements performed at Stanford and at the Maglab) data are reported for the lowest sweep rates (to minimize heating in metallic components of the fridges), and all measurements were performed upon decreasing the magnetic field from fields greater than the critical field.  

To stitch the critical field measurements from both facilities together, measurements were performed in an overlap region (roughly 110 mK to 200 mK). A reduction of approximately 1\% was applied to the critical fields measured at the MagLab High B/T Facility to yield good stitching, which presumably accounts for mismatches in sampling alignment/mounting, slight imperfections in the crystal during shipping/handling that changed its demagnetizing fields, and any other sample/instrument calibration mismatches between the two facilities.

Several attempts were made to obtain reasonable uncertainty values for the critical temperatures and critical fields used to establish the phase boundary as shown in Fig. \ref{phaseDiagram} in the main text. However, the uncertainties obtained arising from the data analysis alone were very small (i.e., the methods of extracting the critical temperature/field were very precise), usually orders of magnitude smaller than the temperature/field points themselves. The temperature and field setpoints were also very stable in the regimes measured, yielding similarly small uncertainties. There are other possible sources of uncertainty, such as systematic uncertainties from the field inhomogeneity of the sample, or random uncertainties from temperature fluctuations from the magnetocaloric effect of TmVO$_4$, among others. For the purposes of obtaining reasonable uncertainties for the fit parameters to the mean-field model of the phase boundary, a heuristic temperature uncertainty of 2\% of the temperature setpoint was used as well as a heuristic magnetic field uncertainty of 10 Oe was used for temperatures above 100 mK and 25 Oe was used for temperatures below 100 mK. It is important to stress that if the very small uncertainties were used for the temperature/field points in the phase boundary, the uncertainties reported for the fit parameters would be unnecessarily large and effectively meaningless. It is noteworthy that the values for the uncertainties only manifest as minor changes in the values of the fit parameters, and the overall effects of the hyperfine interaction on the shape of the phase diagram in TmVO$_4$ reamin consistent.

\section{Hamiltonian calculations}

\subsection{Mean-field theory}\label{supp_mean_field_theory_derivation}
The Hamiltonian in its most general form is:
\begin{equation}
    H = -\frac{1}{2}\sum_{\langle i,j \rangle}J_{ij}^y  {S^y_i}  {S^y_j} -\Gamma^z\sum_i  {S^z_i} - \Gamma^x\sum_i {S^x_i} - \frac{1}{2} \sum_{\langle i,j \rangle} J_{ij}^z  {S^z_i}  {S^z_j} - A^z \sum_i  {I^z_i}  {S^z_i}.
\end{equation}
Here, $J^y$ is the quadrupolar coupling, $\Gamma^z$ is the transverse magnetic field ($\frac{1}{2} g_c \mu_B B_z$), $\Gamma^x$ is the transverse strain ($\eta_1 \varepsilon_1$), $J^z$ is the dipolar coupling, and $A^z$ couples the nuclear spin ($I^z$) to the electronic spin ($S^z$)).

To solve the Hamiltonian, a mean-field approach is used, where each spin is treated as a sum of its mean value and a term corresponding to fluctuations. This means that the first term would become (ignoring terms like the fluctuations squared):

\begin{equation}
    -\frac{1}{2}\sum_{\langle i,j \rangle}J_{ij}^y  {S^y_i}  {S^y_j} = -\frac{1}{2}\sum_{\langle i,j \rangle}J_{ij}^y \big(\langle {S^y_i} \rangle\langle {S^y_j} \rangle + \langle {S^y_i} \rangle\delta {S^y_j} + \delta {S^y_i}\langle {S^y_j} \rangle\big).
\end{equation}
Applying this to the complete Hamiltonian yields:
\begin{equation}
    H \approx H_{mf} = H_{constant} + H_{linear},
\end{equation}
where:
\begin{equation}
    H_{constant} = -\frac{1}{2}\sum_{\langle i,j \rangle}J_{ij}^y \langle {S^y_i} \rangle \langle {S^y_j} \rangle -\Gamma^z\sum_i \langle {S^z_i} \rangle -\Gamma^x\sum_i \langle {S^x_i} \rangle - \frac{1}{2} \sum_{\langle i,j \rangle} J_{ij}^z \langle {S^z_i} \rangle \langle {S^z_j} \rangle - A^z \sum_i \langle {I^z_i} \rangle \langle {S^z_i} \rangle ,
\end{equation}
and:
\begin{equation}
\begin{split}
    H_{linear} = &-\frac{1}{2}\sum_{\langle i,j \rangle}J_{ij}^y \big(\langle {S^y_i} \rangle\delta {S^y_j} + \delta {S^y_i}\langle {S^y_j} \rangle\big) -\Gamma^z\sum_i \delta {S^z_i} -\Gamma^x\sum_i \delta {S^x_i} \\ &- \frac{1}{2} \sum_{\langle i,j \rangle} J_{ij}^z \big(\langle {S^z_i} \rangle\delta {S^z_j} + \delta {S^z_i}\langle {S^z_j} \rangle\big) 
    - A^z \sum_i \big(\langle {I^z_i} \rangle\delta {S^z_i} + \delta {I^z_i}\langle {S^z_i} \rangle\big) .
\end{split}
\end{equation}
Each site can be treated identically, which reduces $H_{linear}$ to:
\begin{equation}
    H_{linear} = -J^y \langle {S^y} \rangle {S^y} -\Gamma^z  {S^z} -\Gamma^x  {S^x} - J^z \langle {S^z} \rangle {S^z} - A^z \big(\langle {I^z} \rangle {S^z} +  {I^z}\langle {S^z} \rangle\big).
\end{equation}
In this four state system, the spin matrices are ($\sigma_{x,y,z}$ are the Pauli matrices and $I_2$ is the $2\times2$ identity):
\begin{align}
    S^{x,y,z} & = \sigma_{x,y,z}\otimes I_2 \\ 
    I^{x,y,z} & = I_2 \otimes \sigma_{x,y,z}.
\end{align}
Therefore, the Hamiltonian can be completely represented as:
\begin{equation}
    H \approx H_{constant} -J^y \langle {S^y} \rangle{S^y} -\Gamma^z {S^z} -\Gamma^x {S^x} - J^z \langle {S^z} \rangle{S^z} - A^z \big(\langle {I^z} \rangle{S^z} + {I^z}\langle {S^z} \rangle\big).
\end{equation}
Excluding the $H_{constant}$ term and rearranging, the Hamiltonian is:
\begin{equation}
    H \approx -J^y \langle {S^y} \rangle{S^y} + \big(-\Gamma^z - J^z \langle {S^z} \rangle -  A^z\langle {I^z} \rangle\big) {S^z} - \Gamma^x S^x - A^z\langle {S^z} \rangle {I^z}.
\end{equation}
For purposes of simplification, the Hamiltonian is written as:
\begin{equation}
    H \approx x{S^x} + y{S^y} + z{S^z} + f{I^z},
\end{equation}
where:
\begin{align}
    x &= -\Gamma^x \langle {S^x} \rangle \\
    y &= -J^y \langle {S^y} \rangle \\
    z &=  -\Gamma^z - J^z \langle {S^z} \rangle - A^z\langle {I^z} \rangle\\
    f &=  -A^z\langle {S^z} \rangle.
\end{align}
The eigenvalues of this Hamiltonian are:
\begin{align}
    \lambda_1 &= -f-\sqrt{x^2 + y^2 + z^2} \\
    \lambda_2 &= f-\sqrt{x^2 + y^2 + z^2} \\
    \lambda_3 &= -f+\sqrt{x^2 + y^2 + z^2} \\
    \lambda_4 &= f+\sqrt{x^2 + y^2 + z^2}.
\end{align}
The partition function is:
\begin{equation}
    Z = \sum_{i = 1}^4 \exp{(-\beta \lambda_i)} = 4 \cosh{(f\beta)}\cosh{(\sqrt{x^2 + y^2 + z^2}\beta)} .
\end{equation}
The expectation value of $S^y$, $S^z$, and $I^z$ are:
\begin{align}
    \langle {S^y} \rangle &= -\frac{ y \tanh{(\beta\sqrt{x^2 + y^2 + z^2} )}}{\sqrt{x^2 + y^2 + z^2}}\\
    \langle {S^z} \rangle &= -\frac{ z \tanh{(\beta\sqrt{x^2 + y^2 + z^2})}}{\sqrt{x^2 + y^2 + z^2}}\\
    \langle {I^z} \rangle &= -\tanh{( \beta f)}.
\end{align}
To solve for the expectation values of each parameter, this system of three implicit equations is solved.

\subsection{Determination of the critical temperature and critical field from susceptibility measurements} \label{supp_determining_critical_tempfield}
\begin{figure}[h]
\centering
  \includegraphics[width=1\linewidth]{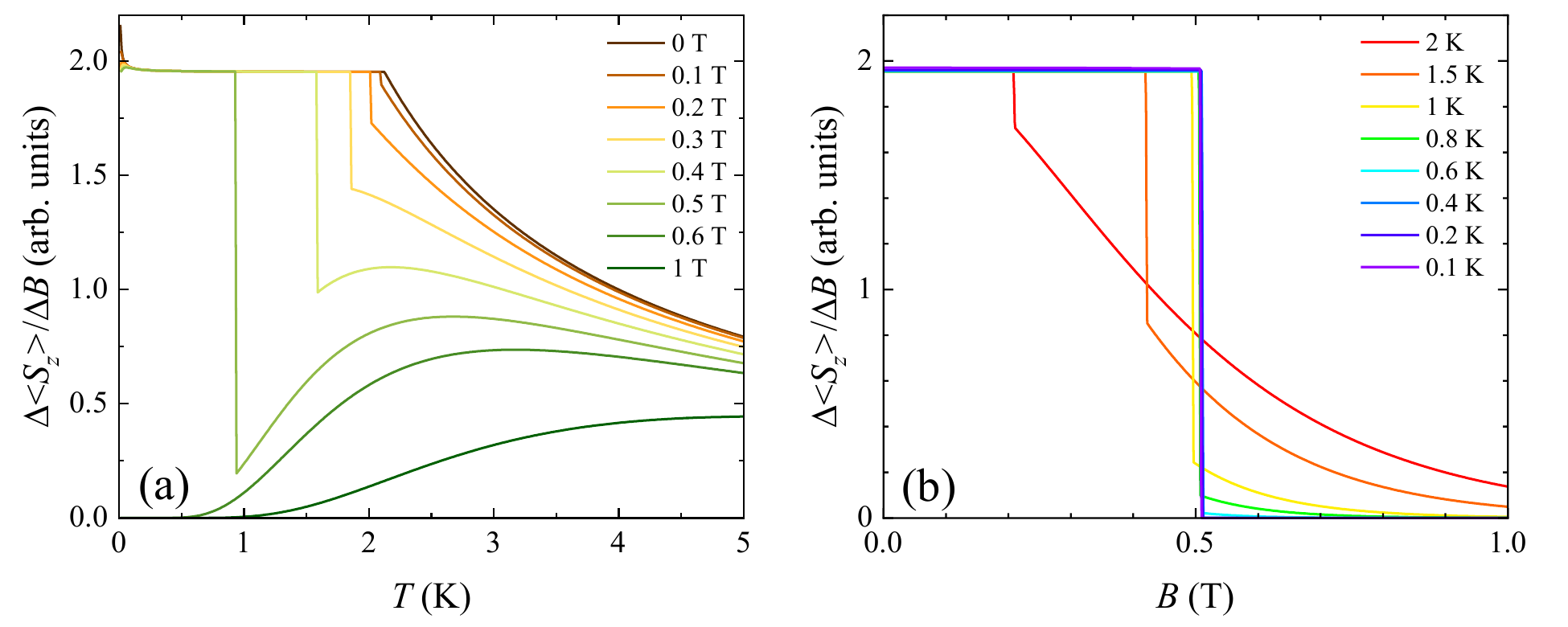}
  \caption{\emph{Calculated susceptibility for the mean-field model using parameters extracted from the fitted phase boundary.} (a) Temperature-dependence of the susceptibility for representative magnetic fields. (b) Field dependence for representative temperatures.}
\label{CalculatedSusceptibility}
\end{figure}

The mean-field model described above readily permits calculation of the longitudinal magnetic susceptibility. Examples are shown in Fig. \ref{CalculatedSusceptibility} as a function of temperature for different fields (panel (a)), and as a function of field for different temperatures (panel (b)). These curves clearly motivate how the phase transition can be determined from the associated susceptibility measurements (Figs. \ref{chiVsT} and \ref{chiVsH} in the main text). In both cases, the actual data are rounded due to field inhomogeneity arising from demagnetization effects, but the criteria defined in the next paragraph mark the transition for the bulk of the material.

Considering first the temperature dependence, it is important to note that for zero applied field the susceptibility exhibits a sharp kink, thus a discontinuous jump in the first derivative. Therefore, for zero-field measurements, the critical temperature is marked by a minimum in the second derivative. However, for non-zero values of the field, the phase transition is marked by a discontinuous jump in the susceptibility, and therefore a minimum in the first derivative. These criteria are used to determine the critical temperature from the data shown in Fig. \ref{chiVsT}. 

\begin{figure}[h]
\centering
  \includegraphics[width=0.6\linewidth]{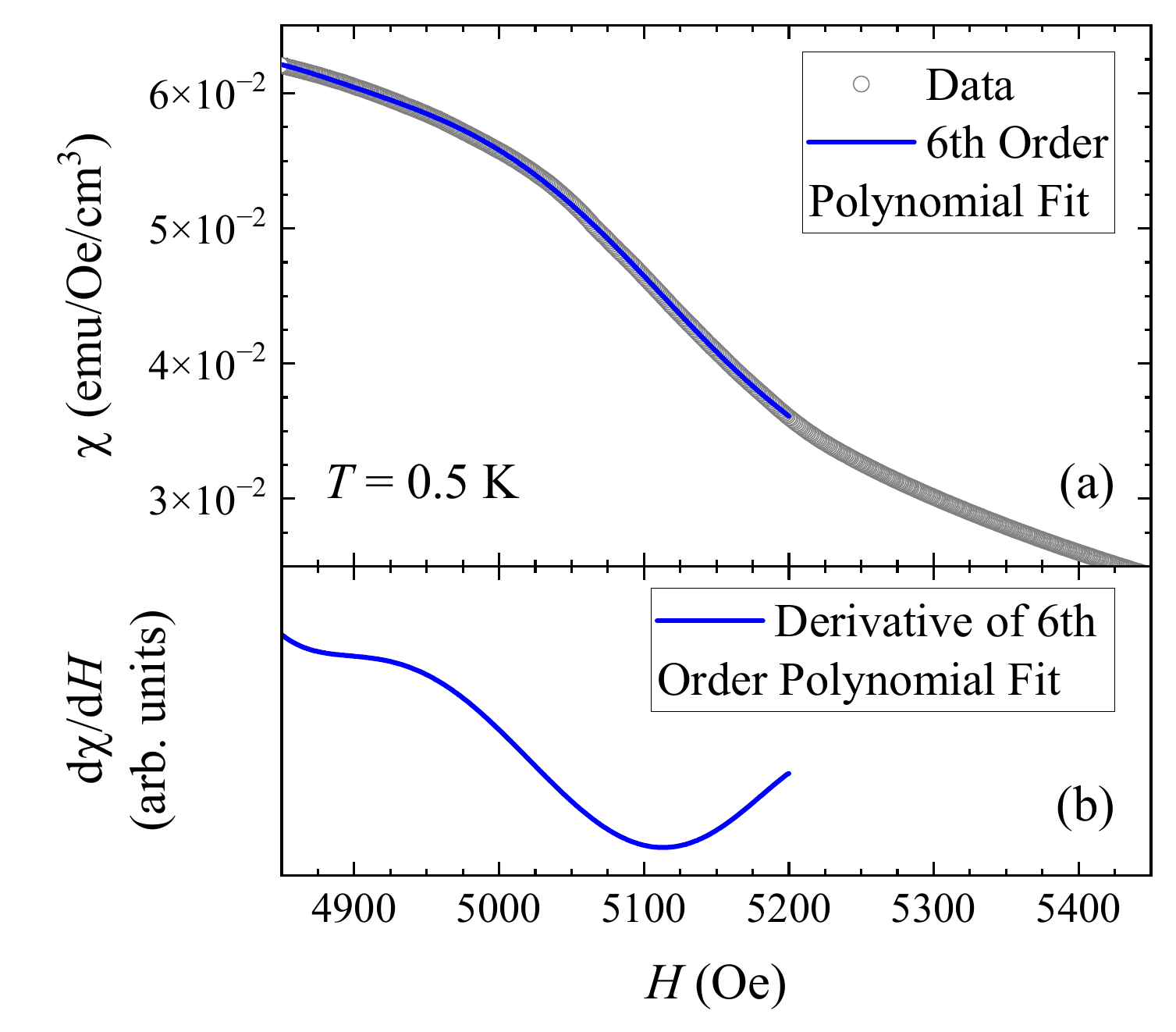}
  \caption{\emph{Representative data (here for a temperature of 0.5 K) showing (a) the susceptibility with 6th order polynomial fit and (b) its first derivative as a function of field.} The minimum of the first derivative of the fit was used to determine critical field values shown in the main text. Only part of the data was fit to a polynomial to allow for better precision over the field range of interest. Windows over which the fit was performed were adjusted based on the temperature, and subsequently the anticipated critical field. }
\label{representative_example}
\end{figure}

Considering the field-dependence, for all temperatures below $T_Q$, the phase transition is marked by a discontinuous jump in the susceptibility, and therefore a minimum in the first derivative.  This criteria is used to determine the critical field from the data shown in Fig. \ref{chiVsH}. As noted in the main text, the jump in the susceptibility is largest at the lowest temperatures, making this the ideal physical quantity to measure in order to determine the critical field at low temperatures. Representative data are shown in Fig. \ref{representative_example}.

\subsection{Entropy landscape as a function of temperature and field} \label{supp_entropy_landscape_full}

\begin{figure}
\centering
  \includegraphics[width=0.8\linewidth]{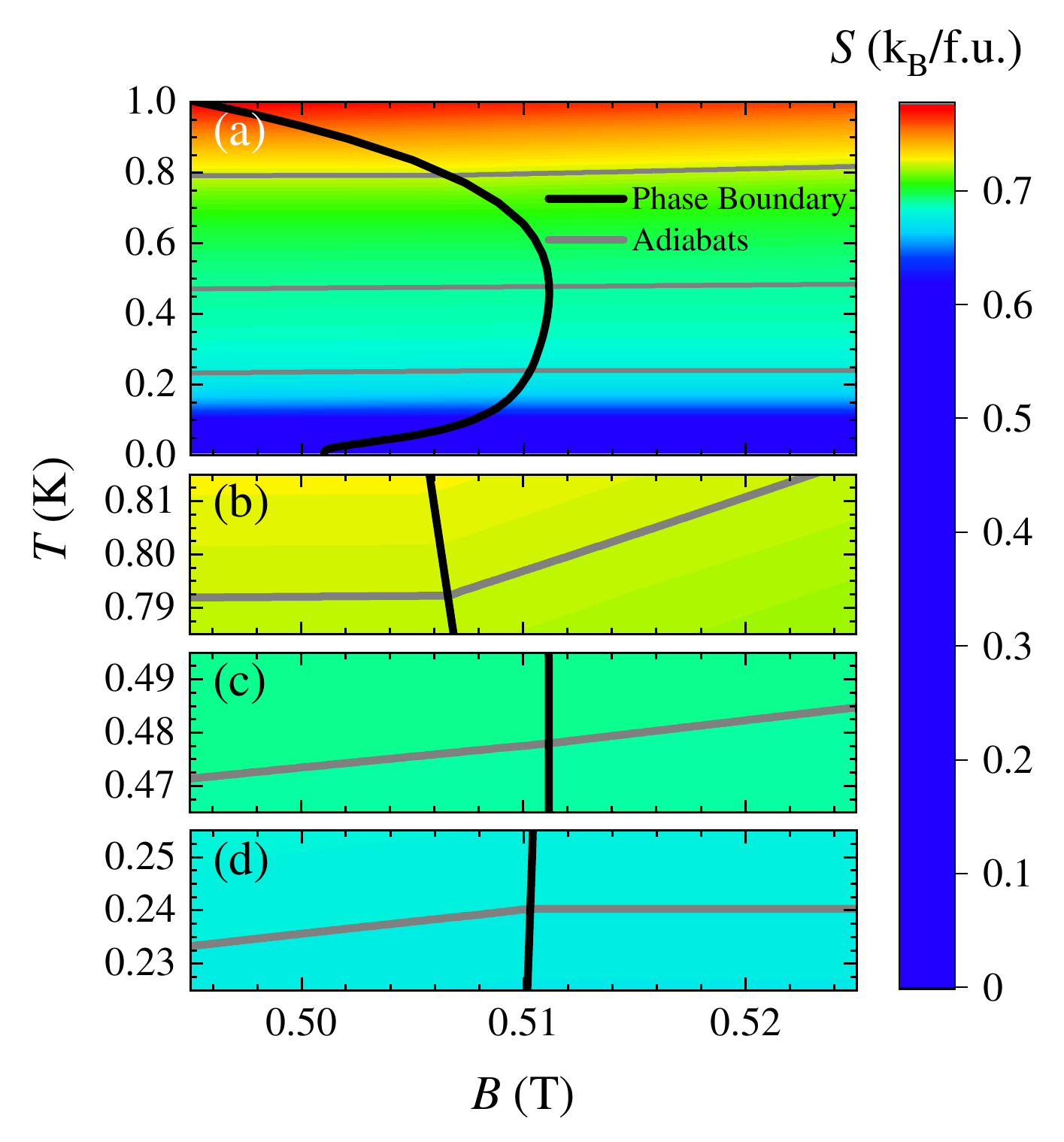}
  \caption{
\emph{Colorscale map showing the calculated total entropy of TmVO$_4$ as a function of temperature and magnetic field, with representative isentropes shown as gray lines.} Panel (b), (c), and (d) are magnified views of the top, middle, and bottom isentropes in panel (a), respectively, shown to reveal changes in the slope upon crossing the ferroquadrupole phase boundary. The phase boundary in all panels is depicted by a solid black line, and the color scale is the calculated entropy expressed per formula unit (i.e., per Tm ion) with a scale that is chosen to most clearly illustrate the changes in entropy in the regimes of interest. Below the crossover temperature of approximately 500 mK, upon reducing the magnetic field the isentropes follow an almost horizontal path in $T$-$B$ space in the paramagnetic phase, but then have a much stronger positive slope in the ordered phase. This effect arises from the hyperfine interaction and is intimately related to the back-bending of the ferroquadrupolar phase boundary (see discussion in main text). Adiabatic demagnetization in this regime would result in a cooling effect. }
\label{entropyLandscape}
\end{figure}

In addition to Fig. \ref{fullEntropyLandscapes} in the main text, it is also helpful to follow the temperature and field-dependence of the calculated entropy very close to the ferroquadrupolar phase boundary in the viscinity of the back-bending as shown in Fig. \ref{entropyLandscape}. Here, panel (a) shows the entropy from zero to 1.0 K for fields ranging from 0.495 T to 0.525 T. Gray lines show representative isentropes in the three distinct regimes discussed in the main text; these three isentropes are then expanded in panels (b) through (d). These show respectively the regime where cooling via electronic demagnetization has the largest effect outside the ordered state; the crossover regime where both effects are operative; and the regime where cooling via nuclear demagnetization inside the ordered phase has the dominant effect.  

\subsection{Nuclear ordering temperature} \label{supp_nucl_ordering_temp_derivation}

To calculate the zero-field, zero-orthogonal strain nuclear critical ordering temperature, it can be noted that the nuclear order happens deep in the ordered state, so $\langle {S^y} \rangle \approx 1$. Because the necessary criterion for order is only that the order parameter become finite, an approximation can be used. $\langle {I^z} \rangle$ can be approximated to be:
\begin{equation}
    \langle {I^z} \rangle \approx A^z \langle {S^z} \rangle \beta.
\end{equation}
Substituting this approximation into the expression for $\langle {S^z} \rangle$ becomes:
\begin{equation}
    \langle {S^z} \rangle = -\frac{( - J^z \langle {S^z} \rangle - A^z\langle {I^z} \rangle) \tanh{(\beta\sqrt{(J^y)^2 + ( - J^z \langle {S^z} \rangle - A^z\langle {I^z} \rangle)^2})}}{\sqrt{(J^y)^2 + (- J^z \langle {S^z} \rangle - A^z\langle {I^z} \rangle)^2}}.
\end{equation}
At low temperatures, terms with $\beta$ dominate. Therefore, $\beta \cdot J^y$ is large, and the $\tanh$ term can be approximated as $1$, leaving the expression:
\begin{equation}
    \langle {S^z} \rangle = -\frac{( - J^z \langle {S^z} \rangle - A^z\langle {I^z} \rangle) }{\sqrt{(J^y)^2 + (- J^z \langle {S^z} \rangle - A^z\langle {I^z} \rangle)^2}} = \frac{  J^z \langle {S^z} \rangle + A^z\langle {I^z} \rangle }{\sqrt{(J^y)^2 + (J^z \langle {S^z} \rangle + A^z\langle {I^z} \rangle)^2}}.
\end{equation}
One can then define a function $f(\langle {S^z} \rangle)$:
\begin{equation}
    f(\langle {S^z} \rangle) = \langle {S^z} \rangle - \frac{  J^z \langle {S^z} \rangle + A^z\langle {I^z} \rangle }{\sqrt{(J^y)^2 + (J^z \langle {S^z} \rangle + A^z\langle {I^z} \rangle)^2}}.
\end{equation}
The value of $\beta$ for which the phase transition occurs is when $f'(0) = 0$, which occurs when:
\begin{equation}
    \beta = \frac{J^y - J^z}{(A^z)^2},
\end{equation}
or at a temperature of:
\begin{equation}
    T = \frac{1}{\beta} = \frac{(A^z)^2}{J^y-J^z},
\end{equation}
which corresponds to approximately 0.96 mK for the values of $A^z$, $J^y$, and $J^z$ determined from fitting the phase boundary:
\begin{align}
    A^z &= 0.0434\ \mathrm{K} \\
    J^y &= 2.12\ \mathrm{K} \\
    J^z &= 0.16\ \mathrm{K}.
\end{align}

\section{Ehrenfest relation} \label{supp_ehrenfest_relation}

Along the phase boundary of a second-order phase transition, the entropy of the two phases are equal at any point, as well as their derivatives:
\begin{align}
    S_{FQ} &= S_{PM} \\
    dS_{FQ} &= dS_{PM}.
\end{align}
Here, $FQ$ denotes the ferroquadrupolar phase and $PM$ denotes the paramagnetic phase. These two expressions can be expanded in the temperature-field plane as:
\begin{align}
    dS_{FQ} &= \frac{\partial S_{FQ}}{\partial H}\Big|_T dH + \frac{\partial S_{FQ}}{\partial T}\Big|_H dT\\
    dS_{PM} &= \frac{\partial S_{PM}}{\partial H}\Big|_T dH + \frac{\partial S_{PM}}{\partial T}\Big|_H dT.
\end{align}
Equating the two expressions and rearranging, while noting that this expression is true for $T=T_Q$ yields:
\begin{equation}
    \frac{d T_Q}{dH} = -\frac{\frac{\partial S_{FQ}}{\partial H}\Big|_T -\frac{\partial S_{PM}}{\partial H}\Big|_T}{\frac{\partial S_{FQ}}{\partial T}\Big|_H - \frac{\partial S_{PM}}{\partial T}\Big|_H}.
\end{equation}
The adiabatic magnetocaloric effect $\frac{d T}{dH}\Big|_S$ in each phase at $T_Q$ is obtained from Eqns. 35 and 36 by setting $dS$ to zero, yielding:
\begin{equation}
    \frac{d T}{dH}\Big|_S = -\frac{\frac{\partial S}{\partial H}\Big|_T}{\frac{\partial S}{\partial T}\Big|_H},
\end{equation}
which rearranges to give:
\begin{equation}
    \frac{\partial S}{\partial H}\Big|_T = -\frac{C_H}{T} \frac{dT}{dH}\Big|_S,
\end{equation}
where the substitution:
\begin{equation}
\frac{\partial S}{\partial T}\Big|_H = \frac{C_H}{T}
\end{equation}
has been made and where appropriate subscripts are used for each phase (FQ and PM) in Eqns. 38-40. 

Substituting for $\frac{\partial S}{\partial H}\Big|_T$ and $\frac{\partial S}{\partial T}\Big|_H$ in Eqn. 37 for each phase yields the final result quoted in the main text:
\begin{equation}
    \frac{d T_Q}{dH} = \frac{\Delta (C_H (\frac{\partial T}{\partial H})_S)}{\Delta C_H},
\end{equation}
where $\Delta C_H = C_H^{FQ}-C_H^{PM}$ etc as defined in the main text.

\section{High temperature magnetic susceptibility measurements} \label{supp_highT_susc_measurements}

\begin{figure}[t]
\centering
  \includegraphics[width=0.8\linewidth]{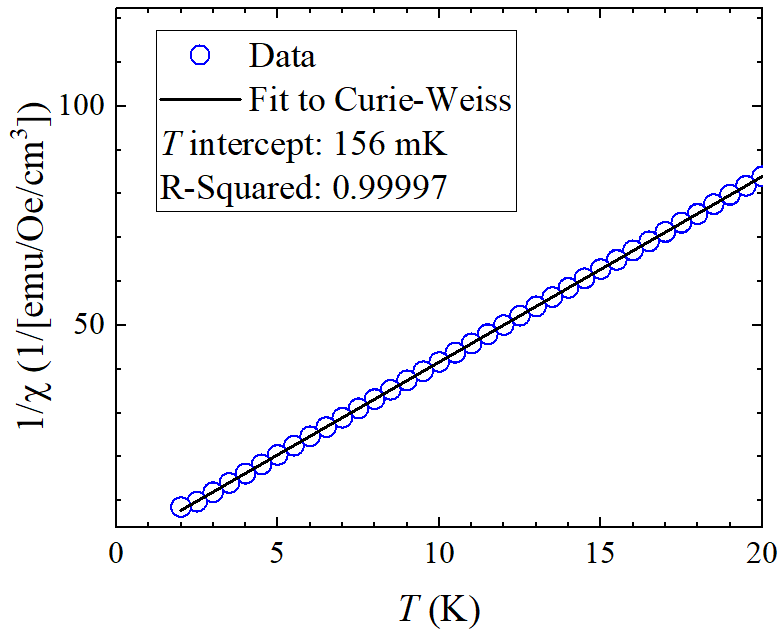}
  \caption{
\emph{High temperature susceptibility data and Curie-Weiss fit.} The reciprocal zero-field magnetic susceptibility data obtained between 2 K and 20 K are fit to an inverse Curie-Weiss form. The fit is excellent and produces a Curie temperature of $\theta = 156$ mK. }
\label{CurieWeissFit}
\end{figure}

The magnetic ordering temperature extracted from magnetic susceptibility measurements and the term $J^z$ in the Hamiltonian are one and the same. For comparison, the magnetic susceptibility measurements between 2 K and 20 K were fit to the reciprocal of a Curie-Weiss form of $T-\theta$ as shown in Fig. \ref{CurieWeissFit}. The fit is very good with an $R^2$ of 0.99997 and yields a Curie temperature of $\theta = 156$ mK, agreeing with that obtained from a fit of the phase diagram in the main text.

\end{document}